\documentclass[authoryear]{elsarticle}
\usepackage{subcaption}
\usepackage{gensymb}
\usepackage{xcolor}
\usepackage[colorlinks]{hyperref}
\usepackage{rotating}
\usepackage{caption}
\newcommand{\e}[1]{\ensuremath{\times 10^{#1}}}
\newcommand{\unit}[1]{\ensuremath{\,\mathrm{#1}}}
\mathchardef\mhyphen="2D
\newcommand{\tf}[1]{\scriptsize{\textbf{#1}}}
\usepackage[autolanguage, np]{numprint}

\graphicspath{{./}{figures/}}

\begin{document}
\begin{frontmatter}

\journal{Icarus}
\title{In-situ detection of Europa's water plumes is harder than previously thought.}

\affiliation[1]{organization={School of Physics and Astronomy, University of Southampton},
            addressline= {University Road}, 
            city={Southampton},
            postcode={SO17 1BJ}, 
            country={UK}}
\affiliation[2]{organization={ESTEC (ESA)},
            addressline={Keplerlaan 1}, 
            city={Noordwijk},
            postcode={2201 AZ}, 
            country={Netherlands}}
\affiliation[3]{organization={Leiden University},
            addressline={Rapenburg 70}, 
            city={Leiden},
            postcode={2311 EZ}, 
            country={Netherlands}}
\affiliation[4]{organization={Space and Planetary Science Centre, Khalifa University},
            addressline={Zone 1}, 
            city={Abu Dhabi},
            country={UAE}}
\affiliation[5]{organization={Department of Mathematics, Khalifa University},
            addressline={Zone 1}, 
            city={Abu Dhabi},
            country={UAE}}
\affiliation[6]{organization={Landessternwarte, Universitat Heidelberg},
            addressline={Konigstuhl}, 
            city={Heidelberg},
            postcode={D 69117}, 
            country={Germany}}
\affiliation[7]{organization={European Southern Observatory},                         addressline = {Karl-Schwarzschild-Str 2},
                city = {Garching},
                postcode = {85748},
                country ={Germany}}
\affiliation[8]{organization={University of Texas at Austin},
            addressline={110 Inner Campus Drive}, 
            city={Austin},
            postcode={78705}, 
            state={Texas},
            country={USA}}
\affiliation[9]{organization={Royal Belgian Institute of Space Aeronomy},
            addressline={Av. Circulaire 3}, 
            city={Brussels},
            postcode={1180}, 
            state={Uccle},
            country={Belgium}}
            
\author[1,2,3]{Rowan Dayton-Oxland\corref{cor1}} 
\cortext[cor1]{Corresponding author.}
\ead{R.A.Dayton-Oxland@soton.ac.uk}
\author[2,4,5]{Hans L. F. Huybrighs} 
\author[6,7]{Thomas O. Winterhalder} 
\author[8,9]{Arnaud Mahieux} 
\author[8]{David Goldstein}

\begin{abstract}
    Europa's subsurface ocean is a potential candidate for life in the outer solar system. It is thought that plumes may exist which eject ocean material out into space, which may be detected by a spacecraft flyby. Previous work on the feasibility of these detections has assumed a collisionless model of the plume particles. New models of the plumes including particle collisions have shown that a shock can develop in the plume interior as rising particles collide with particles falling back to the moon's surface, limiting the plume's altitude. These models also assume a Laval nozzle-like vent which results in a colder plume source temperature than in previous studies, further limiting the plume's extent. We investigate to what degree the limited extent of the shocked plumes reduces the ability of the JUICE spacecraft to detect plume H$_2$O molecules. Results show that the region over Europa's surface within which plumes would be separable from the H$_2$O atmosphere by JUICE (the region of separability) is reduced by up to a half with the collisional model compared to the collisionless model. Putative plume sources which are on the border of the region of separability for the collisionless model cannot be separated from the atmosphere when the shock is considered for a mass flux case of 100\unit{kg\,s^{-1}}. Increasing the flyby altitude by 100\unit{km} such that the spacecraft passes above the shock canopy results in a reduction in region of separability by a third, whilst decreasing the flyby altitude by 100\unit{km} increases the region of separability by the same amount. We recommend flybys pass through or as close to the shock as possible to sample the most high-density region. If the spacecraft flies close to the shock, the structure of the plume could be resolvable using the neutral mass spectrometer on JUICE, allowing us to test models of the plume physics and understand the underlying physics of Europa's plumes. As the altitude of the shock is uncertain and dependent on unpredictable plume parameters, we recommend flybys be lowered where possible to reduce the risk of passing above the shock and losing detection coverage, density and duration.
\end{abstract}


\begin{highlights}
    \item An internal shock caused by particle collisions reduces the radial extent of the model plume, resulting in a decrease in plume density as detected by a simulated flyby by as much as three orders of magnitude.
    \item This can halve the area of the Europan surface over which plumes can be differentiated from the H$_2$O atmosphere (region of separability), making plume detection more difficult.
    \item We recommend Europa flybys pass below the plume canopy altitude, e.g. 300\unit{km}, to maximise the chance of detecting a plume. 
    \item Lowering the altitude of the flyby by 100km can increase the chance of detecting a plume by increasing the region of separability by as much as 32\%.
    \item If the flyby passes close to the plume, a neutral mass spectrometer could resolve the structure of the canopy shock - allowing us to probe the plume physics.
\end{highlights}

\begin{graphicalabstract}
\includegraphics[width=\textwidth]{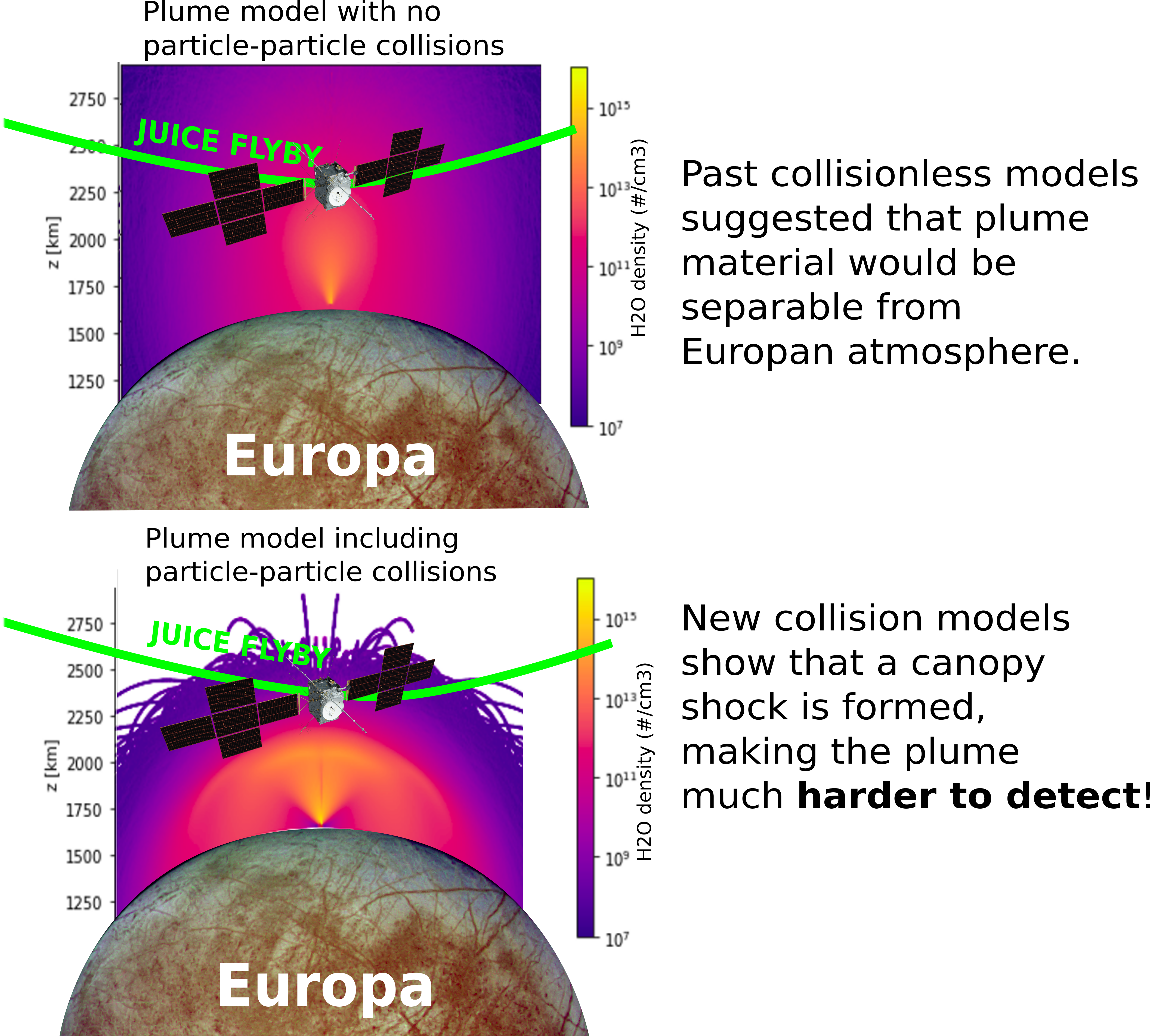}
\end{graphicalabstract}

\end{frontmatter}

\section{Introduction}\label{introduction}
Europa is thought to have a subsurface water ocean with the potential to support life, making it one of the most intriguing places in the solar system to astrobiologists. The physical and chemical characteristics of this ocean are poorly constrained, but it is thought that tidal heating from Jupiter sustains a liquid ocean around 100km deep, which could contain the essential ingredients and energy for life \citep{Nimmo2016,Schubert2004}. Several observation methods have hinted at the possibility of water plumes erupting from the surface, potentially ejecting ocean material into space and providing an opportunity to study its composition \citep{hand2009astrobiology,chyba_europa_2002}. \\

Europa's plumes were first detected by \cite{roth_2014}, who reported evidence of excess UV emission (in the Hydrogen Lyman-$\alpha$ and Oxygen OI 130.4\unit{nm} spectral emission lines) above the southern hemisphere from analysis of Hubble images, consistent with a 200km water plume near Europa's south pole. \cite{sparks_probing_2016,sparks_active_2017,sparks_search_2019} investigated limb-darkening effects in far ultra-violet Hubble images and also found cases of possible plume activity, although this has since been contested by \cite{giono_analysis_2020} who showed that at least one of these detections could also be explained by statistical fluctuations. Recently, studies of Europa's magnetic field and plasma wave observations by \cite{jia_2018} and \cite{arnold_2019} have reported features consistent with interaction between Jupiter's co-rotating plasma and plumes, giving independent evidence to support their existence. Low-albedo geographical regions have also been linked to cryovolcanic activity by \cite{fagents_cryomagmatic_2000}. \cite{paganini_measurement_2020} has shown possible evidence of a plume through IR spectroscopic measurements with the Keck telescope. Recent work by \cite{huybrighs_active_2020} shows that energetic proton depletion during Galileo's flyby of Europa is consistent with charge exchange with a plume, providing evidence from an independent method and data set from previous examples. However, an artefact in the energetic proton data prevents a definitive conclusion on the presence of a plume \citep{Jia2021, Huybrighs2021}. \\

The potential to study the Europan ocean for signs of life has encouraged scientific and public interest in a spacecraft flyby of Europa's plumes. It is proposed that spacecraft flybys could investigate the ocean's composition, search for organic molecules and perhaps sample microbial life directly \citep{lorenz_europa_2016}. In this paper ESA's JUpiter Icy Moons Explorer (JUICE) mission is used as an example mission \citep{grasset_2013}; the spacecraft's Particle Environment Package (PEP) \citep{barabash_particle_2013} has been predicted to be able to detect molecules from the plumes during its two planned flybys of Europa in the early 2030's \citep{huybrighs_2017,tommy_2020}. The spacecraft may be capable of `tasting' trace gases and compounds in the plume using the Neutral Ion Mass spectrometer (NIM) instrument in the PEP package. Europa Clipper's MAss SPectrometer for Planetary EXploration/Europa (MASPEX) instrument is also intended to analyse the composition of Europa’s atmosphere and possible plumes \citep{teolis_plume_2017}, when the spacecraft reaches Europa \citep{brockwell_maspex_2016}.\\

Collisionless models have been used to model plumes on Europa \citep{Southworth2015,lorenz_europa_2016,teolis_plume_2017,huybrighs_2017, Vorburger2021, tommy_2020} and Enceladus \citep{Smith2010, Tenishev2010,Dong2011}. Specifically in \cite{huybrighs_2017} and \cite{tommy_2020}, a simple collisionless model using Monte Carlo particle tracing was used to simulate the trajectories for neutrals and ions under Europa's gravity, Jupiter's electric field, and the conventional electric field, for several possible plume source mass fluxes. These studies show that H$_2$O molecules could be detected by the NIM instrument from a plume anywhere on the surface. However, simulations by \cite{berg_2016} and more recently by \cite{Tseng2022}, which included particle-particle collisions between H$_2$O neutrals resulted in a canopy shock forming within the plume, as particles falling back to the surface collide with rising material. This limits the altitude and extent of the plume. The models presented in \cite{berg_2016} assume a Laval nozzle-like vent which results in a colder plume source temperature than in previous studies, further limiting the plume's extent. The canopy shock and the lower source temperature suggest that in reality a spacecraft's ability to detect plume molecules will be poorer than previously thought.\\

While the structure of Europa's plumes, including the presence of a canopy shock, has not been determined from the currently available observations, the presence of canopy shocks has been observed at Io (e.g. in \cite{Morabito1979,Geissler2008}) and predicted by models (e.g. in \cite{Zhang2003, McDoniel2015, mcdoniel_simulation_2019}). The plumes on Enceladus do not exhibit the canopy shape observed at Io and predicted at Europa. This is because the exhaust velocity of the plumes exceeds the escape velocity of the moon. The plume particles are therefore not strongly affected by gravity and thus the collisions between rising and returning particles are not frequent enough to form the canopy shock (e.g. \cite{yeoh_constraining_2017}). Estimates of plume mass flux range from 5\unit{kg\,s^{-1}} \citep{Southworth2015} to 7000\unit{kg\,s^{-1}} \citep{roth_2014}. \\

In this study the extent to which using a collisional model with a canopy shock reduces the potential of detection compared to a collisionless case is investigated. We determine the effect of particle collisions and the formation of a canopy shock on the region of separability (i.e. the surface area of Europa that can be probed by a flyby), and on the peak density detected by the simulated flyby. The feasibility of detecting particles from previously observed potential plumes by \cite{roth_2014,sparks_probing_2016,sparks_active_2017,sparks_search_2019,jia_2018} and \cite{arnold_2019} are investigated. Flybys at different altitudes are modelled to determine if a lower flyby would significantly improve the area over Europa's surface where plumes would be separable from the atmosphere (the region of separability) and the peak density of particles encountered. \\

\section{Method}\label{method}

The density distribution of the plumes is here simulated using the Direct Simulation Monte Carlo \citep{bird_molecular_1994} code PLANET developed at The University of Texas at Austin. The PLANET code was run with different vent conditions listed in Table \ref{tab:plumeparameters}, varying the mass flux and turning particle collisions on and off. Subsequently, the simulated plume densities are used as an input for simulations of the measurements with the neutral mass spectrometer, NIM, on JUICE.

\subsection{Plume Model}\label{plume_model}

PLANET has been previously used for a wide variety of atmospheric applications, including plume simulations for Europa \citep{berg_2016}, Io \citep{mcdoniel_simulation_2019} and Enceladus \citep{mahieux_parametric_2019,yeoh_constraining_2017}. In this work we simulate the plume H$_2$O molecule number density, velocity, and kinetic- and rotational-temperature fields using the same approach as the one described in \cite{mahieux_parametric_2019}; the reader can find an extensive description and justification of the computation method in that paper. A summary is provided here.\\

\begin{figure}
    \centering
    \includegraphics[width=\textwidth]{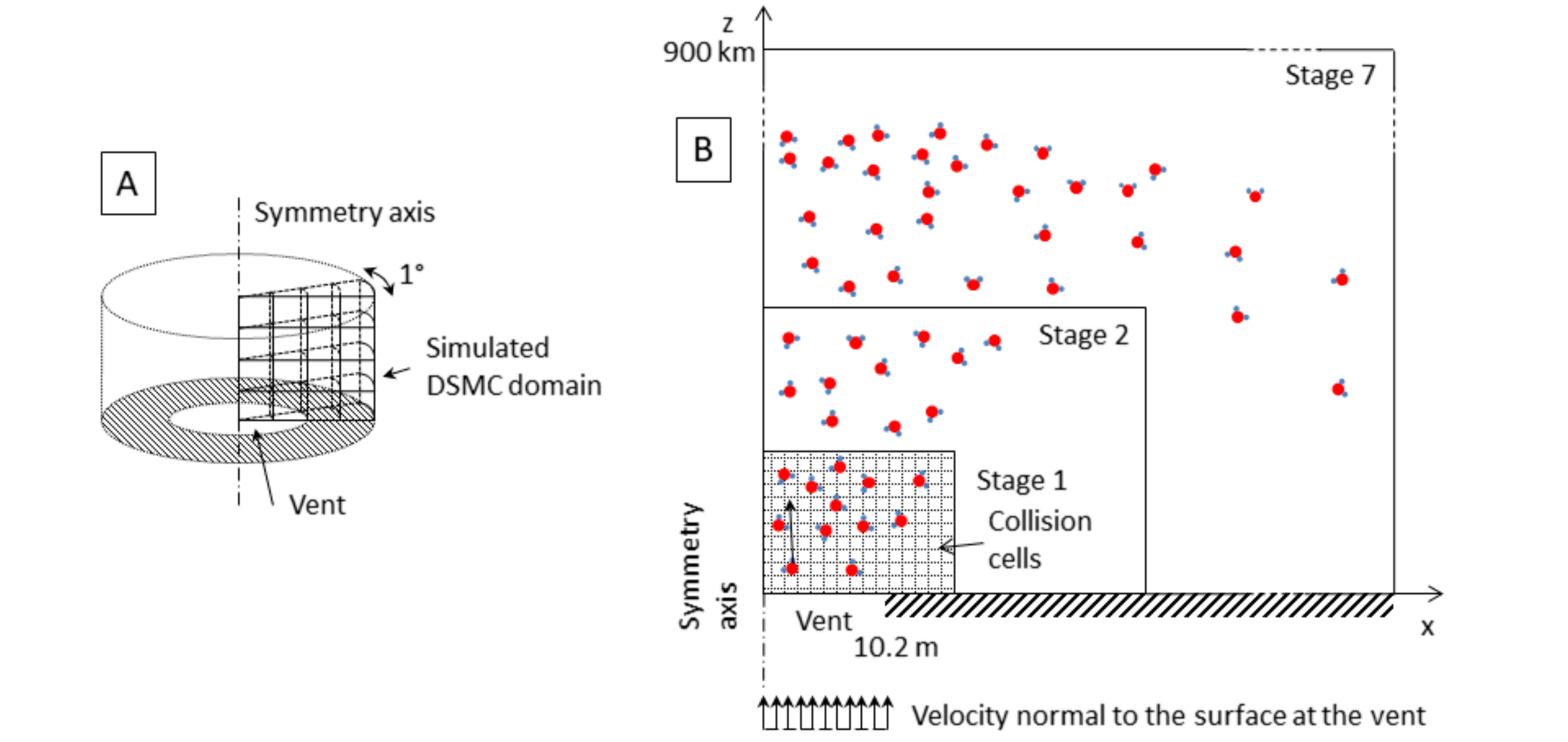}
    \caption{Panel A: DSMC domain geometry. The vent is the white circle at the centre. The grid represents the domain that is simulated using the code PLANET. Panel B: Representation of the staged calculation. The collision cells are shown in Stage 1 only for clarity.}
    \label{fig:simulations}
\end{figure}

A single plume source with a circular vent aperture is considered, pointing normal to the surface of Europa, outgassing a pure water vapour flow. In DSMC, the time-dependent movement and collisions of numerical particles, which represent a large number of real molecules, are computed on a 3D spherical grid. The grid is defined by collision cells, in which the collisions of the particles are computed probabilistically. However, only 1/360 of the flow in a 1-degree wedge needs to be simulated because of the cylindrical symmetry of the problem. The computation is done in stages (see Figure \ref{fig:simulations}), whose sizes increase as a geometric progression, and such that each stage encompasses the previous ones. Thus, the cell size is defined such that it resolves the local flow mean free path. Likewise, the time step needs to resolve the mean collision time.
There are seven stages in the computation. PLANET is run in each stage until steady state is reached, i.e. that the number of numerical particles is constant. When steady state is reached, the flow is sampled over a large number of time steps to reduce the numerical noise, and the numerical particles crossing the right and top boundaries in Figure \ref{fig:simulations} are counted. They are used as initial conditions for the next stage. \\

We note that, in the present DSMC simulations, the particles can only travel from a lower stage to an upper stage. This introduces a small error below when considering Europan plumes, as the particles from the outermost stage that should fall back to stage 6 are not counted. Note, in our simulations the outer stage (stage 7) extends from 40-700\unit{km} from the plume source, and so fully contains the region where the canopy shock forms at an altitude of $\sim400$\unit{km}. As shown in \cite{berg_2016}, this error can be neglected. The stage parameters for the Default case are given in the Appendices, Table \ref{Appendix1}.\\









The collisional model we use has an artefact along the symmetry axis (Figure \ref{fig:simulations} A) (or $z$-axis in Figure \ref{fig:simulations} B), where the density is lower at the centre of the plume than would be expected. This is because there is a statistically small proportion of simulation particles that have the exactly z-aligned velocity needed to populate the region along the central axis. None of the conclusions of this paper are affected by the artefact. The results of Figure 4 and 6 are not affected because we use the maximum density along the entire trajectory and not the value at closest approach (where the artefact could occur). In Figure \ref{fig:densityAT}, panel B, the density at closest approach is an underestimate of less than an order of magnitude compared to panel A. This does not affect the conclusions of the paper.

\subsection{Simulation cases}

For this study, four different DSMC plume simulations were run: a Default case, a second case with 10 times lower mass flux (1.04\unit{kg\,s^{-1}}), a third case with 10 times larger mass flux (104\unit{kg\,s^{-1}}) and a fourth case with no collisions. All cases have the same size vent and the same water vapour speed and temperature at the vent exit. An overview of the simulations is shown in Table \ref{tab:plumeparameters}. The first line refers to the Default case, with collisions turned on. The speed and temperature parameters were taken from \cite{berg_2016}. The stage sizes and grids are the same for all runs. In the Low Flux case, the expanding flow in the neighbourhood of the vent remains collisional only in stage 1 and reaches nearly the same vertical speed as the collisionless case within a few vent diameters above the surface. In the High Flux case, the expanding flow remains collisional up to stage 4. The High Flux plume extends to the same altitude as the Default case. The Default, Low Flux, and High Flux cases plumes all rise to approximately the same altitude. As described in Section \ref{introduction}, estimates from observations show mass fluxes up to $\sim1000$\unit{kg\,s^{-1}}, so the 'high flux' case is high relative to the other models we considered, rather than high relative to the mass fluxes of the observed of plumes. \\

\begin{figure}
    \centering
    \begin{subfigure}{.49\textwidth}
        \includegraphics[width=\textwidth]{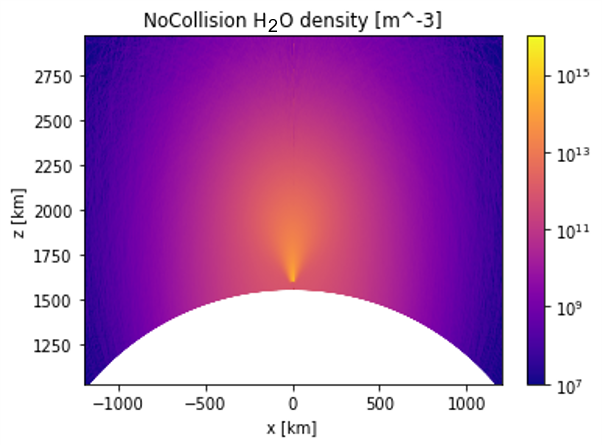}
        \caption{Collision-less}
    \end{subfigure}%
    \begin{subfigure}{.49\textwidth}
        \includegraphics[width=\textwidth]{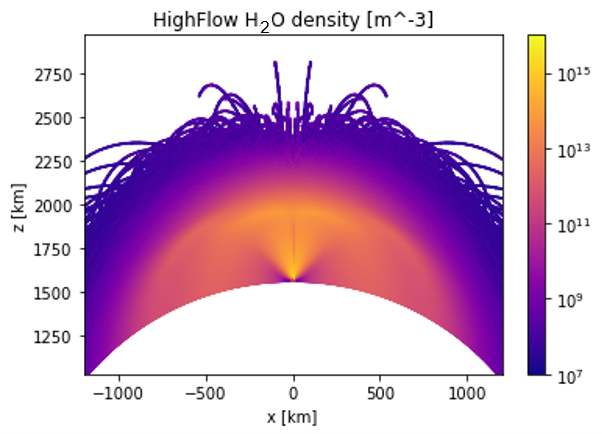}
        \caption{Collisional}
    \end{subfigure}
    \caption{H$_2$O neutral density in the $xz$ plane of two simulated 100\unit{kg\,s^{-1}} plumes. The coordinate system is centred on the centre of Europa with the $z$-axis passing through the plume source. (a) is a collisionless plume showing a large fountain-like shape, (b) is a collisional plume showing the canopy shock. The empty disk segment at the bottom of the figures is the surface of Europa, with the plume source at $z = \mathrm{R}_\mathrm{Europa}$. The canopy shock forms at around $z = 2100$\unit{km} corresponding to an altitude of $\sim 500$\unit{km} above the surface.}
    \label{fig:plume}
\end{figure}

A collisionless plume model was also run. The collisionless plume can be  similarly scaled linearly for different mass fluxes. The collisionless case we expect to be similar to a slightly collisional low-mass flux plume since the flow is so rarefied that collisions between particles hardly occur at any altitude. Therefore, descending particles do not collide with rising particles and a shock wave does not form. Without the canopy shock, the collisionless plume rises higher and spreads more broadly than a collisional, shock-constrained plume would. We expect that the geometric (canopy or no-canopy) nature of Europan plumes is based simply on the mass flow rate and its influence on whether or not a canopy shock forms. Examples of the plume model for a collisional and collisionless 100\unit{kg\,s^{-1}} plume are shown in Figure \ref{fig:plume}. \\

The plume model can be scaled linearly in either the non-shocked (essentially collisionless) regime \citep{huybrighs_2017}, or in the collisional shocked case \citep{berg_2016}. However, during the transition from non-shocked to shocked (roughly between 1 and 10\unit{kg\,s^{-1}}) the scaling of the model is not linear as the structure and height changes from fountain-shaped to the dome-shaped canopy shock. A $\sim1000$\unit{kg\,s^{-1}} plume simulation would yield essentially the same result as a $\sim100$\unit{kg\,s^{-1}} plume, just with a density scaling of $\sim\times10$. \\

\begin{table}[]
    \centering
    \begin{tabular}{p{0.12\linewidth}|p{0.12\linewidth}|p{0.12\linewidth}|p{0.12\linewidth}|p{0.12\linewidth}|p{0.12\linewidth}|p{0.12\linewidth}}
        \textbf{Run Name} & \textbf{Mass Flux in} \unit{kg\,s^{-1}} & \textbf{Vent radius in} m & \textbf{Water vapour exit speed in} \unit{m\,s^{-1}} & \textbf{Water vapour temperature in} K & \textbf{Angle of the Flow to the normal at the vent in} \degree & \textbf{Particle Collisions} \\ \hline
        Default & 10.4 & 10.2 & 902 & 53 & 0 & ON \\
        High Flux & 100.4 & 10.2 & 902 & 53 & 0 & ON \\
        Low Flux & 1.04 & 10.2 & 902 & 53 & 0 & ON \\
        No Collisions & 100.4 & 10.2 & 902 & 53 & 0 & OFF \\
    \end{tabular}
    \caption{Parameters of the vent and of the flow at the vent for the different cases}
    \label{tab:plumeparameters}
\end{table}

The lifetime of a plume particle is sufficiently short such that only a small fraction of the plume particles are ionised by the radiation environment (including e.g. solar UV-C and the Jovian plasma torus). We therefore assume the ion component does not affect the outcome of this study and it is neglected \citep{huybrighs_2017}.\\

\subsection{Instrument Simulation}\label{instrument_simulation}
The example JUICE mission flybys are the baseline 3.0 flybys\footnote{https://www.cosmos.esa.int/web/spice/spice-for-juice} originally planned to take place in October 2030. At closest approach the spacecraft will have an altitude of $\sim400$\unit{km}. Note that since this study is primarily illustrative and the flybys are still planned to pass over each hemisphere with a closest approach at 400 km, the conclusions of this paper are unaffected by the updates to the flybys thereafter. In this study the southern flyby is taken as a reference trajectory, as it passes over the southern hemisphere where most of the putative plume locations lie. The JUICE spacecraft's PEP package contains the NIM instrument; a high resolution time-of-flight mass spectrometer, capable of detecting low energy ($<$10\unit{eV}) neutrals and ions. PEP also contains the ion mass spectrometer, Jovian Dynamics and Composition Analyzer (JDC), which covers the 1\unit{eV} - 41\unit{keV} range. \\

A 1D model of Europa's atmosphere is assumed, with a constant water density of 10$^4$\unit{cm^{-3}}, which is based on densities above 100\unit{km} in model D in \cite{Shematovich2005}, which is the same atmospheric model as used in \cite{huybrighs_2017} and \cite{tommy_2020}. The \cite{Shematovich2005} model is a 1D collisional Monte-Carlo model that includes sublimation and sputtering as sources of H$_2$O.\\

Since we are only looking at altitudes at or above 300\unit{km} we ignore the variations in atmospheric density closer to the surface. This assumption and simplification of the atmosphere is described in \cite{Shematovich2005}; a summary of the main arguments is as follows. Firstly lateral density gradients have been predicted and observed between the day and night side of Europa \citep{teolis_plume_2017,PLAINAKI2010385,Roth2021} as H$_2$O is sublimated from the surface during the day and re-adsorbed at night. 
These modelled and observed asymmetries are hemispheric and so are at a much larger scale than the local density increase caused by the plume, meaning that plumes can be separated from this lateral asymmetry. The atmospheric simplification also ignores secondary effects playing a role in the distribution of plume particles caused by interaction with the surface, atmosphere, and radiation environment such as sputtering/resputtering, bouncing, adsorption/desorption and photon or electron-impact ionisation, modelled for example in \cite{teolis_plume_2017}. Furthermore, these secondary plume features are ignored as they are less dense than the primary erupted particles. 
Since this paper focuses specifically on the effects of collisions in the plumes, we do not attempt to simulate all secondary effects. \\

In this paper a plume detection is defined as the plume signal exceeding NIM's noise level. The noise level is determined from instrumental effects and background radiation represented as a density of 7\unit{cm^{-3}} following \cite{huybrighs_2017}. The density of trace gases, organic molecules, and H$_2$O molecules must surpass this noise limit to be detected by NIM. To separate the plume from the background water atmosphere we require the plume H$_2$O density to exceed the background atmospheric H$_2$O density (at an altitude of $\geq$100\unit{km}) of $10^4$\unit{cm^{-3}}, and define this point as the \textit{atmospheric threshold} \citep{Shematovich2005}. Trace gases and organic molecules are separable from the atmosphere providing they exceed both the instrument noise limit and their species' own atmospheric background density.\\

\begin{figure}
    \centering
    \includegraphics[width=0.9\textwidth]{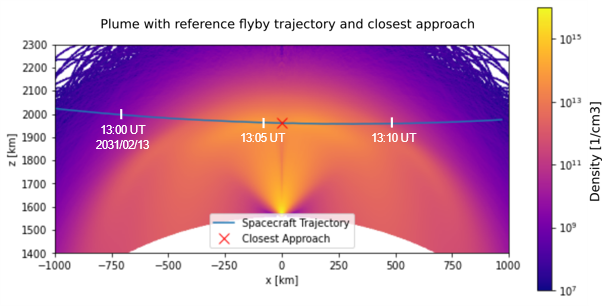}
    \caption{JUICE trajectory through the model 100\unit{kg\,s^{-1}} collisional plume, with closest approach at an altitude of 400\unit{km} (at $z=~2000\unit{km}$). The section of the spacecraft trajectory that we see here covers approximately 15 minutes, and approximate time ticks are shown in white.}
    \label{fig:traj}
\end{figure}

The 2D plume models were converted into a 3D map of H$_2$O neutral density in Europa's vicinity by rotating the single slice by 360\degree{} via an interpolation algorithm. The density of particles encountered along the spacecraft trajectory, as shown in Figure \ref{fig:traj}, was calculated using the same software tools as \cite{tommy_2020}. \\

\section{Results and Discussion}\label{res-dis}

\subsection{Detected density reduced and concentrated around closest approach}\label{densityreduced}
\begin{figure}
    \centering
    \includegraphics[width=\textwidth]{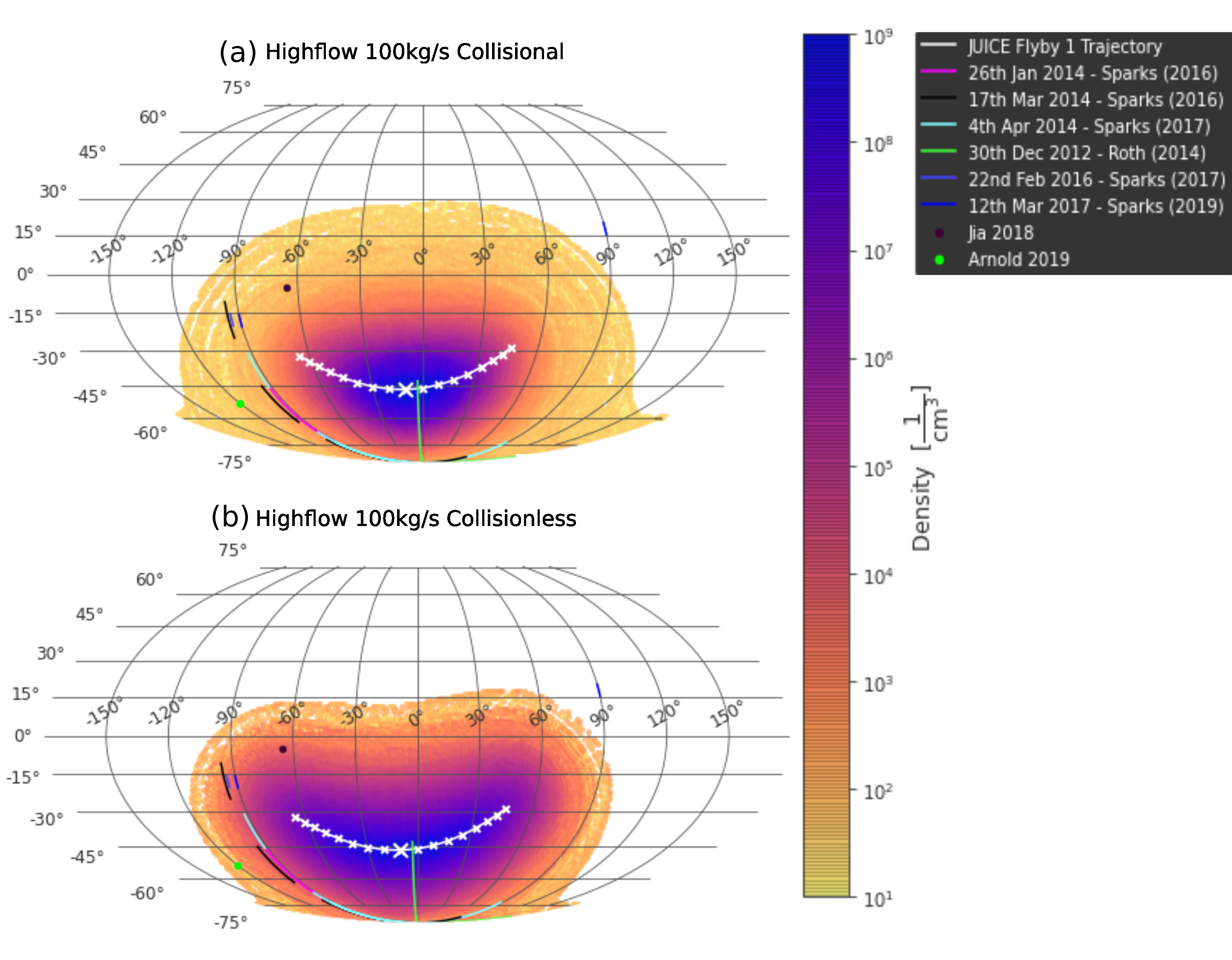}
\caption[short]{\label{fig:plainmaps}Maps illustrating the peak H$_2$O density detectable at the spacecraft trajectory as a function of plume location over the surface of Europa. For a plume located at any point on the surface, the colour at that point corresponds to the peak detected density at the flyby from that plume. The collisionless (a) and collisional (b) models with a mass flux from the source of 100\unit{kg\,s^{-1}} are compared. Approximate locations of putative plumes from the literature are plotted and labelled in the legend. The white background on Europa's surface indicates the regions where the density is less than $10^1\unit{cm^{-3}}$. This plot illustrates that the density detected by the spacecraft for a plume not located at the closest approach is orders of magnitude lower when a collisional model is used, see section \ref{densityreduced}.}
\end{figure}

The ability of the spacecraft to detect plumes over the Europan surface is displayed as detected density maps, showing the peak density detected along the spacecraft trajectory at the instrument as a function of plume location on the map. The maps are generated by considering a plume located at any location on the surface, and determining the peak density by a simulated flyby through the resulting 3D H$_2$O density map of Europan space for each possible plume location. To compare the collisional and collisionless plume cases, the two detected density maps shown in Figure \ref{fig:plainmaps} are produced.\\ 

The collisional model results in a reduced peak detected density, with increasing effect the further away from the location of the closest approach. For example, by inspecting Figure \ref{fig:plainmaps}, a collisional 100\unit{kg\,s^{-1}} model plume located at the position of the putative plume detected in \cite{jia_2018} would lead to a maximum detected density of $10^{1}$\unit{cm^{-3}} encountered by the instrument during flyby, while a collisionless model would suggest a detected density of $10^4$\unit{cm^{-3}}, a difference of three orders of magnitude. For plumes not located directly below the closest approach, the flyby will detect orders of magnitude lower H$_2$O molecule density than was previously predicted from collisionless models. Therefore, plume models that do not include collisions such as the feasibility studies in \cite{huybrighs_2017} and \cite{tommy_2020} lead to over-predictions of the plume H$_2$O density for mass fluxes above 1\unit{kg\,s^{-1}}. \cite{teolis_plume_2017} also use a collisionless model for a plume with a mass flux of 500\unit{kg\,s^{-1}}. Our models predict a strong canopy shock effect at this mass flux, and as such the predicted H$_2$O density by \cite{teolis_plume_2017} from a plume along a spacecraft flyby trajectory is an over-prediction, for flybys with altitudes of several 100\unit{km} that will likely pass over the canopy shock.\\

\begin{figure}
\centering
\begin{subfigure}{.49\textwidth}
    \centering
    \includegraphics[width=\textwidth]{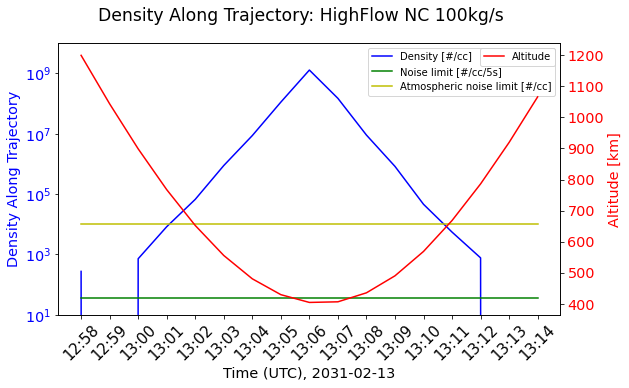}
    \caption{No collisions : 100 kg/s}
\end{subfigure}%
\begin{subfigure}{.49\textwidth}
    \centering
    \includegraphics[width=\textwidth]{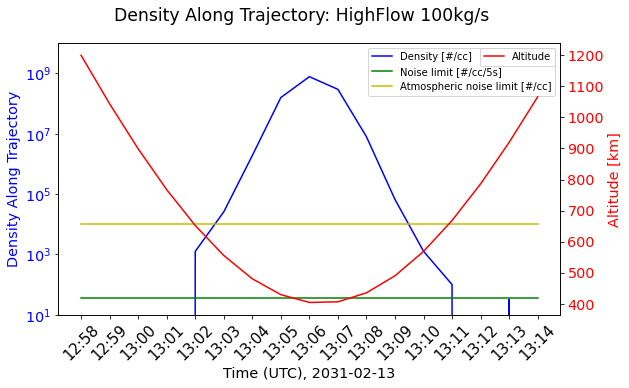}
    \caption{With collisions : 100 kg/s}
\end{subfigure}
\caption[short]{\label{fig:densityAT}Variation of H$_2$0 density with time during the flyby, with closest approach at 400km. The plot shows the density of H$_2$O molecules (blue) detected by the spacecraft in time as it flies through a plume located directly under the point of closest approach, with the atmospheric limit (the approximate water density from $\sim100$\unit{km} to at least 600\unit{km} in the atmosphere from \cite{Shematovich2005}) and NIM noise limit indicated. The red line gives the spacecraft's altitude above the surface in km, the green line gives the NIM instrument noise limit in particles per cubic centimetre; the yellow line indicates the atmospheric H$_2$O density below which plume H$_2$O molecules are not discernible. The density detected by the spacecraft drops off 23\% more quickly with altitude for a collisional plume, see sections \ref{densityreduced} and \ref{plumetime}.}
\end{figure}

The highest density ($>10^6$\unit{cm^{-3}}) region is more restricted to the area of the surface around the closest approach (Figure \ref{fig:plainmaps}), indicating that the relationship between density detected and instrument altitude is stronger in the collisional case. Figure \ref{fig:densityAT} shows how the density changes with altitude as the spacecraft follows its trajectory in time. The gradient of the change of density with altitude is steeper in the collisional case as expected. The density scale factor is defined as the change in altitude at which the density drops by a factor of $e$. For the 100\unit{kg\,s^{-1}} mass flux plume, the density scale factor is $\sim$36\unit{km} using a collisionless model, and is reduced to just $\sim$28\unit{km} for a collisional model. This represents a reduction of 23\%, indicating that the density drops off significantly more quickly when the limiting effect of the shock is considered.\\

These two effects of the collisional model, the reduction in peak detected density and density scale factor, hold for each of the three mass fluxes studied, but is strongest for the high mass 100\unit{kg\,s^{-1}} plume, see Appendices, Figure \ref{fig:appendix_plainmaps}.\\

\subsection{Limitation for plume observation time}\label{plumetime}
The plume observation time is defined as the period during the flyby when the plume density exceeds the atmospheric threshold. For a 100\unit{kg\,s^{-1}} plume located directly below the closest approach, the plume observation time for a collisionless model is 10 minutes. This is reduced to 6 minutes for a collisional plume, a reduction of 40\%. This is a result of the altitude limiting effect of the canopy shock, as there is a shorter period where the spacecraft is at low enough altitude to detect plume H$_2$O molecules at a density above the atmospheric threshold. \\

JUICE's NIM instrument will only be able to make measurements during the approach to Europa. When JUICE enters the outbound phase of the flyby, incoming neutral gas will be blocked from the NIM instrument entrance by the body of the spacecraft \citep{Galli2022}. Applying the halved NIM time to our results, the plume time is reduced to 5 minutes for a collisionless model, and only 3 minutes for the collisional model. Recall that this is for a plume located below the closest approach and represents a best case scenario; the plume time will be even shorter for plumes not in this ideal location. This paper does not fully explore the impact of this restriction on NIM, for example by calculating the detection from only the first half of the flyby. We consider an optimal flyby to illustrate the potential of the instruments and mission. Note that the halved instrument time is a restriction for JUICE specifically and does not necessarily apply to other spacecraft such as Europa Clipper, to which this study is also relevant. \\

The plasma instruments such as the JDC and Energetic Neutral atom imagers in the PEP package do not have the same limitation and will make measurements during both phases. JDC may detect H$_2$O$^+$ pickup ions \citep{huybrighs_2017} created by electron impact ionisation of plume neutral H$_2$O. We expect that the collisional model plume will also limit the physical extent of the region from which pickup ions can be generated as it follows the density distribution of the neutral component, and therefore would also cause a reduction in plume detection time for JDC. \\

\subsection{Collisional model gives reduced region of separability}\label{reducedros}
\begin{figure}
\centering
\includegraphics[width=\textwidth]{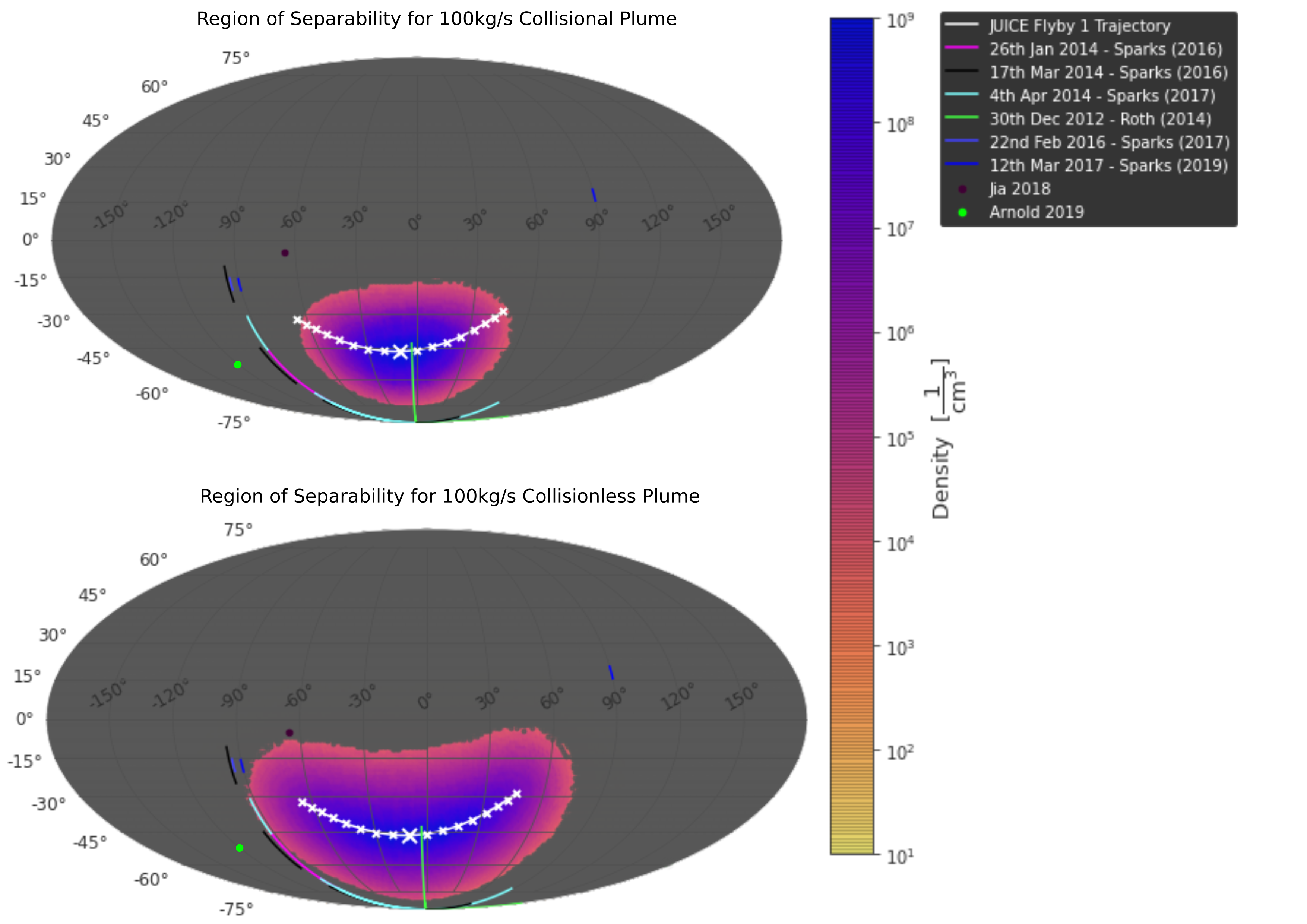}
\caption[short]{\label{fig:atmosmaps}Maps illustrating the H$_2$O molecule density detected in the region of separability, defined as the region in which the detected H$_2$O density exceeds the atmospheric threshold of $10^4$ molecules per \unit{cm^{3}}. These maps show the area of the surface of Europa over which JUICE could separate a plume from the moon's atmosphere (colour), where the greyed out area is everywhere a plume would not be separable by a this flyby. Approximate locations of putative plumes from the literature are plotted on the surface and labelled in the legend. The region of separability is reduced in the collisionless case to about a half of the collisional case, see section \ref{reducedros}.}
\end{figure}

The region of separability can be defined as the area over the surface of Europa where the H$_2$O density from a model plume (in said area) detected by the flyby exceeds the atmospheric threshold and is therefore identifiable as separate from the moon's atmosphere. Maps showing the region of separability considering the atmospheric threshold are shown in Figure \ref{fig:atmosmaps}.\\

Comparing the collisional and collisionless cases there is a reduction in the area of the region of separability (note that the difference in the extent of the simulated particles, visible in Figure \ref{fig:plume}, does not affect the size of the region of separability). In the high mass flux case of 100\unit{kg\:s^{-1}}, the area of the region of separability is reduced by 43\% for the collisional case compared to the collisionless case. Table \ref{tab:massflux_compare} shows the area of the region of separability for each case in \unit{km^{2}}. These data show the region of separability is reduced in each case, and that higher mass flux plumes suffer greater area reduction. For the 1\unit{kg\,s^{-1}} mass flux plume the area is reduced only by 3\%. Hence, at low mass fluxes the detected density is not as greatly affected by the presence of the shock; this is a result of the shock effect being less strong and limiting at low mass flux (see Section \ref{method}). Therefore, a collisionless model as used in \cite{huybrighs_2017} and \cite{tommy_2020} is a useful approximation for modelling plumes with mass fluxes  $\leq1$\unit{kg\,s^{-1}}. Note 1\unit{kg\,s^{-1}} plumes are detectable by JUICE according to \cite{huybrighs_2017, tommy_2020}. See Appendices, Figure \ref{fig:appendix_ros} for region of separability plots for each mass flux. \\

Due to a lack of measurements, uncertainties exist in the atmospheric profile, so better understanding of the atmospheric profile is needed to separate plumes from the atmosphere with high accuracy.\\

\begin{table}
    \centering 
    \begin{tabular}{c|c|c|c}
         \textbf{Mass Flux} & \multicolumn{2}{c|}{\textbf{Region of separability as \%}} & \textbf{Ratio} \\
         \textbf{in \unit{kg\,s^{-1}}} & \multicolumn{2}{c|}{\textbf{of Europan surface}} & \textbf{$\frac{\mathrm{Collisional}}{\mathrm{Collisionless}}$} \\
         & \textbf{Collisional} & \textbf{Collisionless} & \\
         \hline
         100                           & 9.34\%                & 16.50\%                             & 0.57                     \\
         10                            & 7.07\%                & 11.53\%                             & 0.61                     \\
         1                             & 7.17\%                & 7.40\%                        & 0.97                     \\
    \end{tabular}
    \caption{Comparison of the region of separability for a collisional and collisionless model at different plume mass fluxes, where the region of separability is defined as the surface area where the spacecraft is capable of differentiating plume water molecules from the atmospheric water molecules. In the 100\unit{kg\,s^{-1}} case the region of separability is reduced to 57\% of the collisionless model, see section \ref{reducedros}. In the 1\unit{kg\,s^{-1}} case the collisional and collisionless models agree with only 3\% difference.}
    \label{tab:massflux_compare}
\end{table}

\subsection{Implications for detecting putative plumes}\label{putativepl}
To illustrate the implications of the reduction in area of the region of separability, the possibility of detecting putative plumes from the literature is presented in Table \ref{tab:putative}. Considering the case of a 100\unit{kg\,s^{-1}} plume, this table shows whether the putative plumes lie within, on the border, or outwith the region of separability. Putative plumes that lie touching the border of this region are marked as marginal.\\

\begin{table}[ht]
    \centering
    \begin{tabular}{c|c|c|c|c|c|c|c}
        Plume sources               & Collisionless         & Collisional         \\
        \hline
        \cite{sparks_probing_2016}(A) & \colorbox{yellow}{m}  & \colorbox{red}{n}   \\
        \cite{sparks_probing_2016}(B) & \colorbox{yellow}{m}  & \colorbox{red}{n}   \\
        \cite{sparks_active_2017}(A)  & \colorbox{yellow}{m}  & \colorbox{red}{n}   \\
        \cite{roth_2014}            & \colorbox{green}{y}   & \colorbox{green}{y} \\
        \cite{sparks_active_2017}(B)  & \colorbox{red}{n}     & \colorbox{red}{n}   \\
        \cite{sparks_search_2019}   & \colorbox{red}{n}     & \colorbox{red}{n}   \\
        \cite{jia_2018}             & \colorbox{yellow}{m}  & \colorbox{red}{n}   \\
        \cite{arnold_2019}          & \colorbox{red}{n}     & \colorbox{red}{n}     \\
    \end{tabular}
    \caption{\colorbox{green}{y} Separated from atmosphere ; \colorbox{yellow}{m} marginally separated ; \colorbox{red}{n} not separable.\\
    This table illustrates how plumes which were considered separable using a collisionless model are not separable when a collisional model is used, for 100\unit{kg\,s^{-1}} plumes being measured during the southern flyby over the southern hemisphere, see section \ref{putativepl}.}
    \label{tab:putative}
\end{table}

There are putative plumes that would be considered to have marginal possibility of separation using a collisionless model, which are not detectable when a collisional model is used. This is an illustrative example of the concrete effect collisional models have on whether these putative plumes are detectable.\\

\subsection{Lowering altitude of flyby increases region of separability}\label{section:altitude}
Trajectories at closest approach altitudes of 300\unit{km} and 500\unit{km} were tested in addition to the planned JUICE flyby trajectory at 400\unit{km}. This set of trajectories coincidentally aligns with the spacecraft passing under, over, and through the top of the canopy shock. Atmospheric models from \cite{SHEMATOVICH2005480} show that the H$_2$O density in Europa's atmosphere increases below 300\unit{km}. We do not consider altitudes below 300\unit{km} in our analysis, so it is therefore not necessary to consider this atmospheric density increase. \\

\begin{figure}
    \centering
    \includegraphics[width=0.8\textwidth]{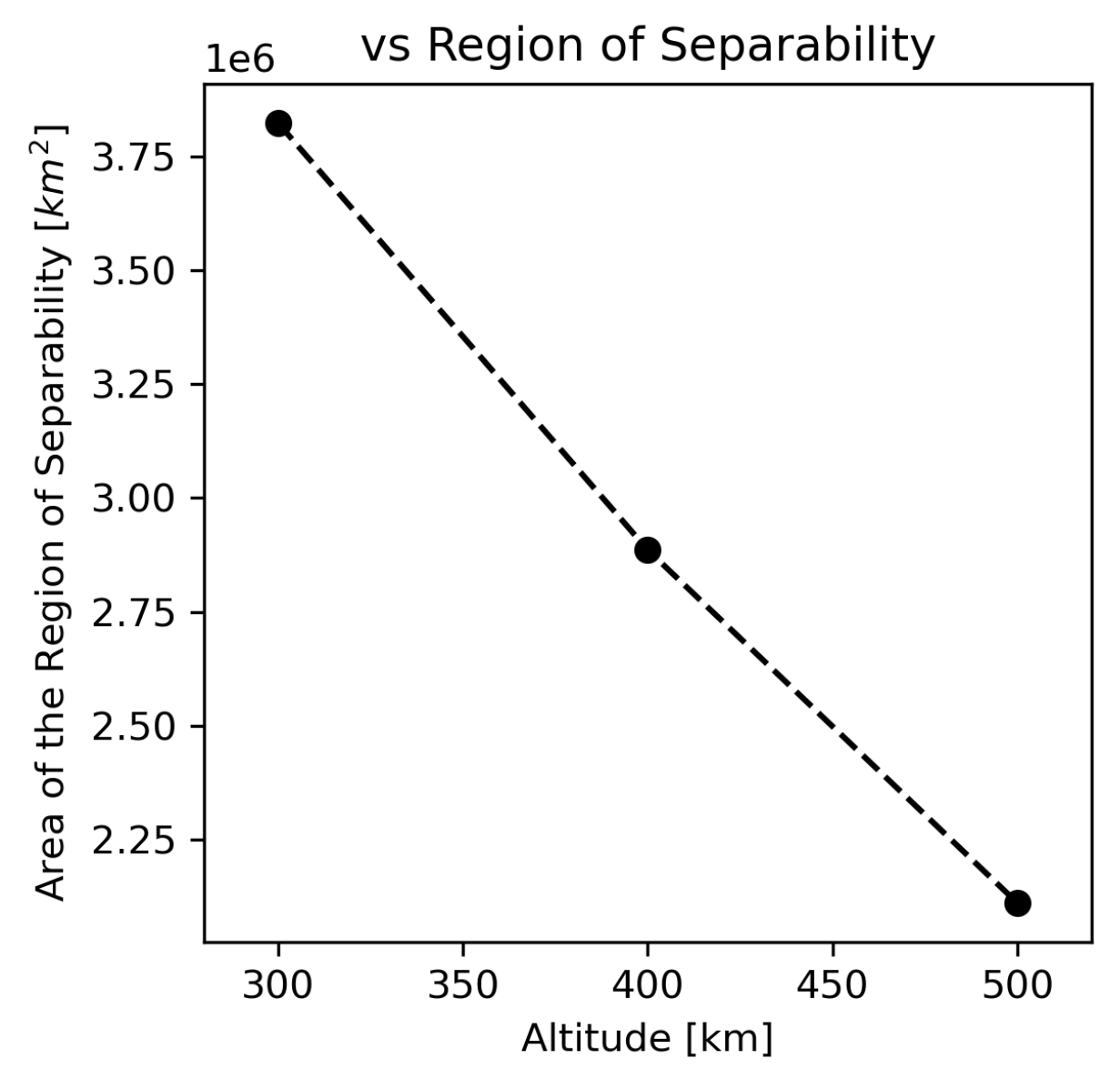}
    \caption{Region of separability i.e. area over the surface of Europa over which a plume would be separable from the Europan atmosphere, plotted against the altitude of the closest approach for three modelled spacecraft flybys. See section \ref{section:altitude}}
    \label{fig:ROSvsALT}
\end{figure}

Table \ref{tab:altitude_compare} and Figure \ref{fig:ROSvsALT} show the result of different trajectory altitudes on the region of separability, see Figure \ref{appendix_altitude} in the Appendices. When the spacecraft flies lower, under the shock canopy altitude, the region of separability is increased by 32\% compared to the 400\unit{km} trajectory. Similarly, when the altitude is increased and the spacecraft passes above the shock canopy the region of separability is reduced by 27\%. The density along the trajectory is affected by an order of magnitude between the 400 and 500\unit{km} cases, and the width of the density-time peak is reduced indicating a shorter time period where detections can be made. Wherein during a 300\unit{km} flyby the spacecraft may detect plumes within a 14\unit{minute} window, this is reduced to 9\unit{minutes} for a 500\unit{km} flyby. \\

\begin{table}
    \centering
    \begin{tabular}{c|c|c}
         \textbf{Altitude at closest}     & \textbf{Region of Separability as a}         & \textbf{Ratio to}   \\
         \textbf{approach in \unit{km}}   & \textbf{fraction of Europa's surface area}   & \textbf{400km case} \\
         \hline
        300                               & 12.4\%              &  1.32       \\
        400                               & 9.3\%               &  1          \\
        500                               & 6.8\%               &  0.73       \\
    \end{tabular}
    \caption{Comparison of the region of separability at different altitudes, based on a collisional 100\unit{kg\,s^{-1}} case. Increasing or decreasing the altitude by 100km affects the region of separability by roughly 30\% each way, see section \ref{section:altitude}.}
    \label{tab:altitude_compare}
\end{table}

The height of the canopy shock of potential plumes is currently poorly constrained, as previous plume detections have all had high uncertainties associated with the size of the plume. Simulations for different source mass fluxes, different vent geometry, or different temperature and velocity parameters suggest a height somewhere in the 100's of kilometres range. In the simulations studied in this paper, the spacecraft trajectory at 400km is sufficient to pass through the plume shock, but this only represents one possible set of plume parameters.\\

\subsection{Effect of low temperature plume compared to previous studies}

Comparing the models presented in this paper to those in \cite{tommy_2020} and \cite{huybrighs_2017} the collisionless model in this work (e.g. Figure \ref{fig:atmosmaps}) predicts a smaller region of separability for the same mass flux. This implies that the model plume in this work has a smaller physical extent to that used in \cite{tommy_2020,huybrighs_2017}. Because this discrepancy in plume extent is visible for even low mass flux and collisionless models where there is no canopy shock, and because the average velocity in this work is higher (902\unit{m\,s^{-1}}) than in \citep{huybrighs_2017, tommy_2020} (460\unit{m\,s^{-1}}) we attribute the difference to our plume source temperature (53\unit{K}), which is colder than in previous studies (230\unit{K} \citep{huybrighs_2017, tommy_2020} and 150\unit{K} \cite{ teolis_plume_2017}). At cooler source temperatures, plume molecules ejected have a lower velocity spread as per their Maxwell-Boltzmann distribution, and thus fewer of them reach distances further away from the source, reducing the extent of the plume. \\

The low plume temperature (53\unit{K}) is due to the following physical argument, as shown in \cite{berg_dsmc_2016}. According to \cite{berg_dsmc_2016}, plume material leaving the vent will be subject to the Laval nozzle effect. This means that the material will be cooled as it erupts out into space, dependent on the vent properties. The plume simulation in this study has a source temperature of 53\unit{K}, based on physical arguments considering a high speed plume starting at 230\unit{K} in the vent, then being cooled by the nozzle effect \citep{berg_2016} to 53\unit{K}. Note that the temperature and exit velocity used in our simulations is the coldest case put forward by \cite{berg_2016} i.e. a worst case scenario in terms of detectability.\\

Plumes may be colder than expected in previous feasibility studies, making them more difficult to detect by spacecraft flyby. Plume temperature should be considered by JUICE and other Europa in-situ missions, as less energetic plumes will have a smaller physical extent and therefore be less likely to be detected by spacecraft flyby.\\

\subsection{Discerning plume structure}
The plume canopy shock is a large structure in comparison with the path of the spacecraft, and may intersect the trajectory for around 500\unit{km} (in $x$, see Figure \ref{fig:traj}) before and after the closest approach. This translates to several minutes of plume observation time (see subsection \ref{plumetime}) which is much longer than the NIM integration time (5 seconds), giving good temporal (and therefore spatial) resolution of samples of plume material. Presence of a plume would dominate the H$_2$O density detected during in-situ detections during the flyby for a plume located near the closest approach, meaning that any changes in density throughout the plume would be due to the structure of the plume itself rather than the background atmosphere. This suggests that the structure of the plume could be discernible from the density of plume molecules encountered during the flyby. Measurements of the structure of the plume would provide an empirical constraint on plume models, by which we can investigate the underlying physics and typical characteristics of Europan plumes. This would allow us to investigate yet unknown properties of plumes such as the temperature, exit velocity, mass flux, and vent characteristics.
\\

\section{Recommendations for future missions}\label{section:implications}
We recommend that missions such as JUICE and Europa Clipper consider low flyby altitudes for Europa. For the best chance of detection of any plume, including those not yet detected, the flyby should take place below the shock altitude e.g. 300\unit{km}, to give the highest chance of flying through or below some part of the canopy and making a positive plume detection (i.e. maximising the region of separability). However, if a specific active plume is targeted, we recommend to target the shock itself as it is an extended, high density region which is difficult to miss (as opposed to the dense region directly above the source, which is a fraction of the size of the shock and therefore easier to miss). \\

Putative plume locations from the literature are present primarily around Europa's southern hemisphere, close to the track of the southern flyby. The putative plume detected in \cite{roth_2014} shows the greatest chance of being detected by JUICE as it is positioned close to the planned closest approach for the southern flyby. The estimated mass flux of this plume is $1-7\e{3}$\unit{kg\,s^{-1}}; our models predict that this mass flux would easily lead to a canopy shock which would constrain the extent of the plume and therefore the region of separability and plume density (at altitudes above the canopy). \\

The JUICE flyby over the southern hemisphere is the better candidate (of the two planned flybys) for plume detection as it covers more putative plume candidates, see \cite{tommy_2020}. If lowering any of the flybys is possible the southern flyby should be chosen, as flying low over putative plumes increases the chance of a positive detection.\\

\cite{tommy_2020} showed that when using a collisionless model, lowering the flyby altitude would not result in a significant increase in the region of separability, so the benefit of a lowered altitude flyby was only in the increase in density of the detected plume H$_2$O molecules. However, this study using a collisional model shows that the reduced extent of the plume due to the canopy shock can lead to a large (43\%) reduction in the area of the region of separability, which can only be offset by lowering the flyby altitude by 100\unit{km}. Hence, the case study presented in \cite{tommy_2020} must be re-assessed. In contrast to their results, we find that lowering the flyby trajectory would increase the likelihood of separating putative plumes from the atmosphere, when we account for particle collisions. \\

\cite{teolis_plume_2017} predicts the H$_2$O atmosphere will be suppressed on the night side due to surface condensation, but enhanced on the day side due to sublimation, which was observed in \cite{Roth2021}. As the night-side will therefore have a lower atmospheric water density at the flyby altitude, the atmospheric threshold will be lower leading to a larger region of separability. Plumes would therefore be easier to detect by spacecraft flyby from their H$_2$O signature on the night-side of Europa. The JUICE mission’s proposed imaging for Europa requires optimal imaging illumination conditions i.e. a $\beta$-angle (angle between the orbital plane and the sun) of $\geq$ 50\unit{\degree}. This means it is likely that most of the JUICE trajectory will be on the dayside – however the proposed orbits by Europa clipper should sample the nightside thoroughly \citep{JUICE, Bayer2019}.\\ 

The canopy altitude of the plume will depend on various parameters, such as the mass flux, temperature of the flow, exit speed, and spreading angle at the vent, as well as vent geometry. All of these are poorly constrained on Europa. Full exploration of the possible canopy altitudes is beyond the scope of this paper, though a full parameter sweep would be a useful study. Though the plume height is difficult to constrain, the recommendations presented in this work would apply to any plume of 100s\unit{km} scale - the most plausible case. Models presented by \cite{Zhang2003} and \cite{McDoniel2015} explore the dependence of canopy height on vent temperature and exit velocity, but without further constraints on the plume physics it is not currently possible to give a height range beyond the 100s\unit{km} scale discussed here.

\section{Conclusions}\label{conclusions}
The predicted probability of detecting plume H$_2$O molecules is reduced when considering a collisional plume model compared to collisionless, for mass fluxes exceeding 1\unit{kg\,s^{-1}}. For the low mass flux plume (1\unit{kg\,s^{-1}} or lower) the collisional model approximates the collisionless model. In the high mass case for a plume with flux of 100\unit{kg\,s^{-1}}, the useful region of separability is reduced by as much as half when compared to the collisionless model, and the peak particle density detected at closest approach can be reduced by as much as an order of magnitude. Previously suggested putative plume locations that were thought to be detectable with the southern JUICE flyby may lie outwith the region of separability when collisions are considered. \\

Model-based predictions for detecting Europa's plumes from a spacecraft flyby must consider particle collisions, as the results obtained using a collisional model show differences as large as order of magnitude lower density, 23\% steeper altitude density gradient, and 43\% reduced region of separability in the high mass flux (100\unit{kg\,s^{-1}}) case. The nozzle effect results in much colder plume temperatures than have been used in previous feasibility studies. Since this limits the extent of the plume, it should also be considered in future plume modelling - particularly for plume detection where the physical extent of the plume is vital for a positive detection/separation. \\

If a flyby is low enough to pass through or under the plume shock, for a model 100\unit{kg\,s^{-1}} case, the region of separability is improved by 32\%. Conversely, increasing the flyby altitude to 500km reduces the region of separability by 27\%. For a spacecraft aiming to detect plume molecules, a lower altitude is recommended to improve the likelihood of passing through plumes originating from the widest possible area of the surface. The alternative scenario where the spacecraft passes over the shock altitude, results in a region of separability reduced by almost a third, and peak density reduced by an order of magnitude.\\

A flyby passing through the plume shock could give a plume observation window of 3 minutes (for a model including particle collisions). Compared to JUICE's neutral mass spectrometer (NIM) observing cadence of 5 seconds, we see that the structure of the plume could be resolved, allowing us to probe the underlying physics of Europa's plumes. Resolving the structure of the plume would allow us to compare the observation to plume models and infer plume parameters such as mass flux, vent characteristics, or plume temperature.

\section*{Declaration of competing interest.}
The authors declare that they have no known competing financial interests or personal relationships that could have appeared to
influence the work reported in this paper.

\section*{Data availability}
Data will be made available on request.

\section*{Acknowledgements}
We thank the reviewers for their effort in providing their useful and insightful comments on the manuscript. This work was carried out during the Leiden/ESA Astrophysics Program for Summer students (LEAPS) 2020 hosted by the European Space Research and Technology Centre (ESTEC) - European Space Agency (ESA) and Leiden Observatory. R. Dayton-Oxland is supported by the University of Southampton and INSPIRE Doctoral Training Program; this work was supported by the Natural Environmental Research Council [grant number NE/S007210/1]. H. Huybrighs gratefully acknowledges the  support from Khalifa University’s Space and Planetary Science Center under grant N. KU-SPSC-8474000336, the ESA research fellowship and the International Space Science Institute (ISSI) visiting scientist program. T. O. Winterhalder was supported by the University of Heidelberg and the European Southern Observatory. D. B. Goldstein and A. Mahieux were supported by the SSW NASA grant with award number 80NSSC21K016. Plume computations were done at the Texas Advanced Computing Center.\\

\bibliographystyle{elsarticle-harv} 
\bibliography{main}

\section*{Appendices}

\label{Appendix1}
\begin{sidewaystable}[ht]
    \centering
    \begin{tabular}{c|c|c|c|c|c|c|c|c|c|c|c}
         &  & \tf{\# vert.} & \tf{\# horiz.} & \tf{Stg vt.} & \tf{Stg hz.} & \tf{Time} & \tf{Mean coll.} & \tf{Cell vt.} & \tf{Cell hz.} & \tf{Mean free} & \\ 
         \tf{Stg} & \tf{$f_{num}$} & \tf{cells} & \tf{cells} & \tf{size [m]} & \tf{size [m]} & \tf{step [s]}  &  \tf{time [s]} & \tf{size [m]} &  \tf{size [m]} &  \tf{path [m]} &  \tf{$Kn_{grad}$}\\ \hline
        1 & $9\e{16}$ & 144 & 6 912 & 44.9 & 44.9 & $10^{-4}$ & $5\e{-4}$ & 0.31 & 0.006 & 0.09 & 0.06 \\
        2 & $2.3\e{17}$ & 144 & 6 912 & 112.7 & 78.8 & $2\e{-4}$ & 0.0023 & 0.78 & 0.011 & 0.31 & 0.2  \\
        3 & $6.7\e{17}$ & 240 & 11 520 & 333.4 & 189.2 & $5\e{-4}$ & 0.0088 & 1.39 & 0.016 & 0.95 & 0.5  \\
        4 & $2.3\e{18}$ & 288 & 13 824 & 1 163.7 & 604.3 & 0.002 & 0.095 & 4.04 & 0.043 & 7.63 & 1.7  \\
        5 & $9.6\e{18}$ & 384 & 18 432 & 4 790.8 & 2 417.9 & 0.009 & 2.3 & 12.48 & 0.13 & 110.9 & 6.9  \\
        6 & $4.7\e{19}$ & 528 & 25 344 & 23 262.4 & 11 653.7 & 0.05 & 82.3 & 22.07 & 0.46 & 2495 & 36.7  \\
        7 & $1.8\e{21}$ & 1 008 & 48 384 & 900 000 & 900 000 & 2 & $4\e6$ & 892.86 & 18.6 & $6.2\e7$ & 3.2 \\
    \end{tabular}
    \captionsetup{singlelinecheck=off}
    \caption[.]{Mean values of DSMC parameters in each stage (stg) for the Default case: $f_{num}$, number of vertical (vt.) and horizontal (hz.) cells, horizontal and vertical size of each stage, time step, spatially averaged mean collision time, cell vertical size, spatially averaged mean free path, and spatially averaged Knudsen number $Kn_{grad}$.\\
    The computational values are mean values over the whole stage. The computational time step resolves the mean collision in all stages. Furthermore, the horizontal cell sizes always resolve the mean free path. Note that the vertical cell size does not resolve the mean free path in stages 1 and 2. However, the flow is close to continuum conditions in those lowest stages, as the Knudsen number based on the density gradient is very small.
    $$Kn_{grad} = \lambda_{MFP}\cdot\frac{\|\nabla m_{H_2O}\|}{m_{H_2O}}$$
    where $\lambda_{MFP}$ is the mean free path assuming local thermodynamic equilibrium and $m_{H_2O}$ is the local water vapour mass density. It varies between $2.42\e{-5}$ and 0.1 in stage 1 and between 0.04 and 0.5 in stage 2.}
    \label{tab:dsmcparameters}
\end{sidewaystable}
    
\begin{figure}[h]
\centering
\includegraphics[width=0.6\paperwidth]{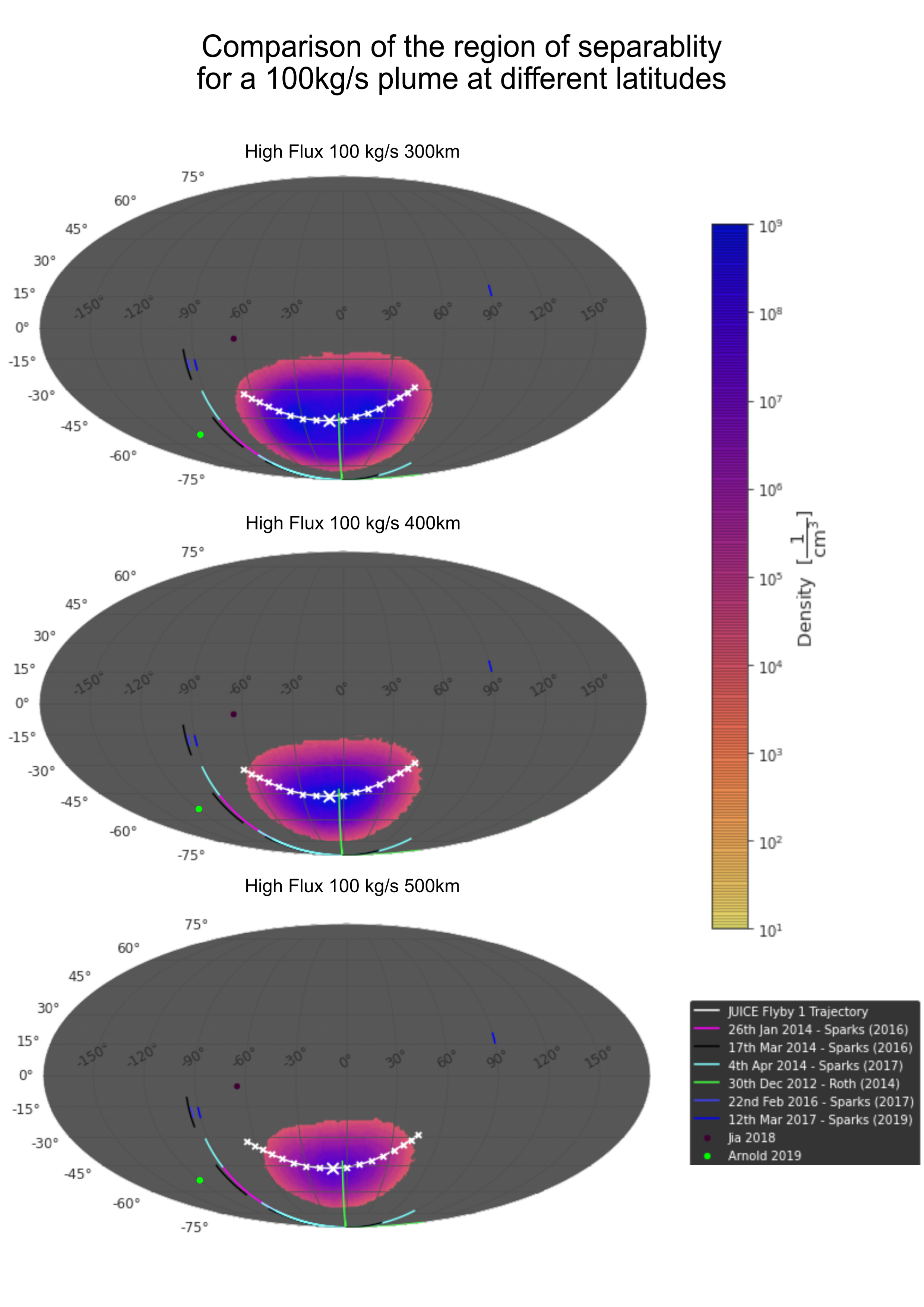}
\caption[short]{\label{appendix_altitude} Comparison of the region of separability at difference trajectory altitudes at closest approach. The region of separability for the 500km case (bottom) above the plume is reduced by a third compared to the default 400km (middle), where in the 300\unit{km} case (top) the region of separability is improved by a third.}
\end{figure}

\begin{figure}
    \centering
    \includegraphics[width=0.65\paperwidth]{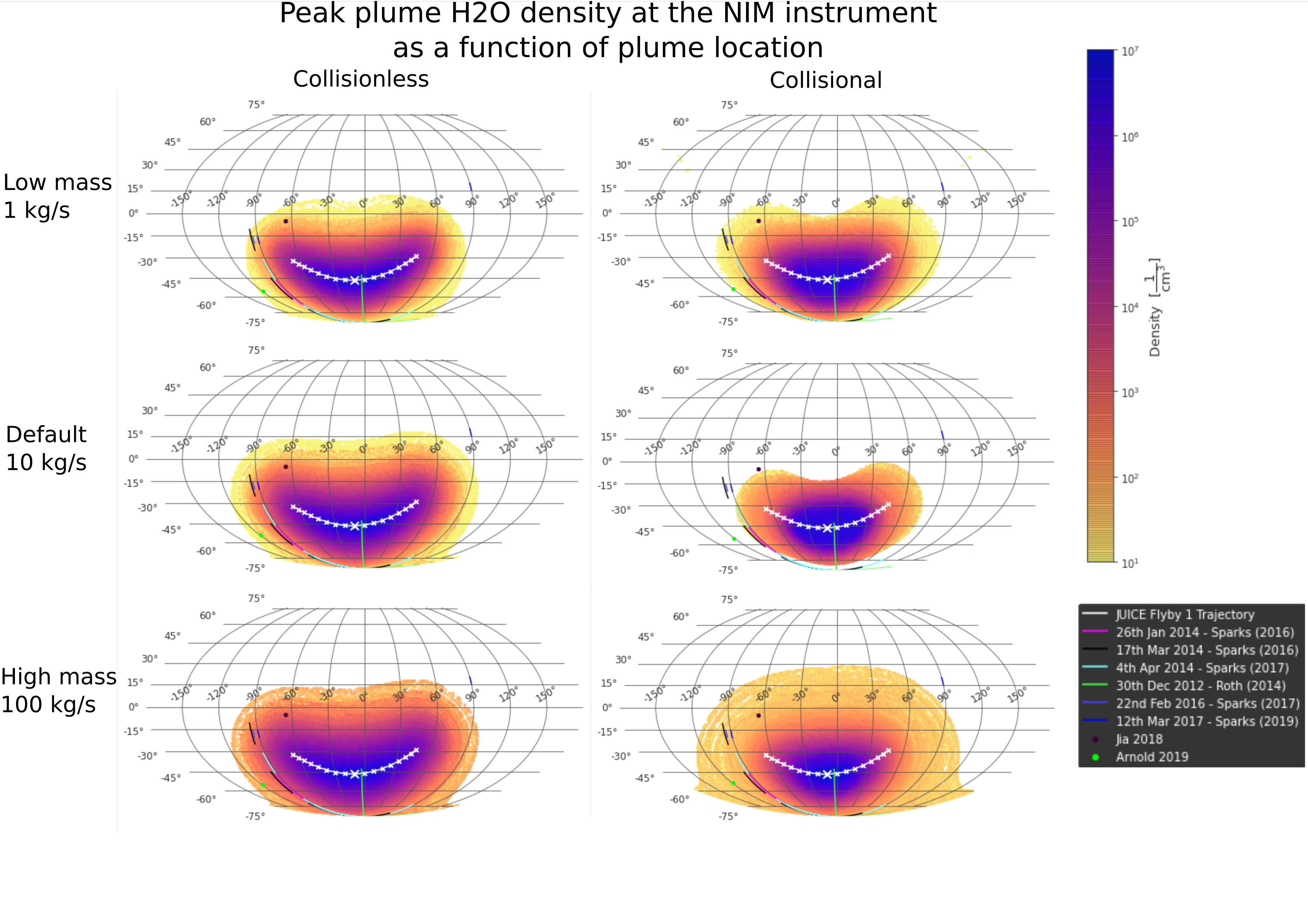}
    \caption{Plot of the peak H$_2$O density detected by the NIM instrument during the flyby as a function of the plume location on the surface, for different plume models consisting of the collisional and collisionless model of a low mass 1\unit{kg\,s^{-1}}, 10\unit{kg\,s^{-1}}, and 100\unit{kg\,s^{-1}}. For a low mass plume the effect of the particle-collisions is lesser, so the collisional map is more similar to the collisionless case than in the high mass plume case, where the collisional and collisionless plume density maps are drastically different. This figure illustrates how higher mass flux plumes are more affected by the addition of particle collisions as the canopy shock effect is stronger.} 
    \label{fig:appendix_plainmaps}
\end{figure}

\begin{figure}
    \centering
    \includegraphics[width=0.65\paperwidth]{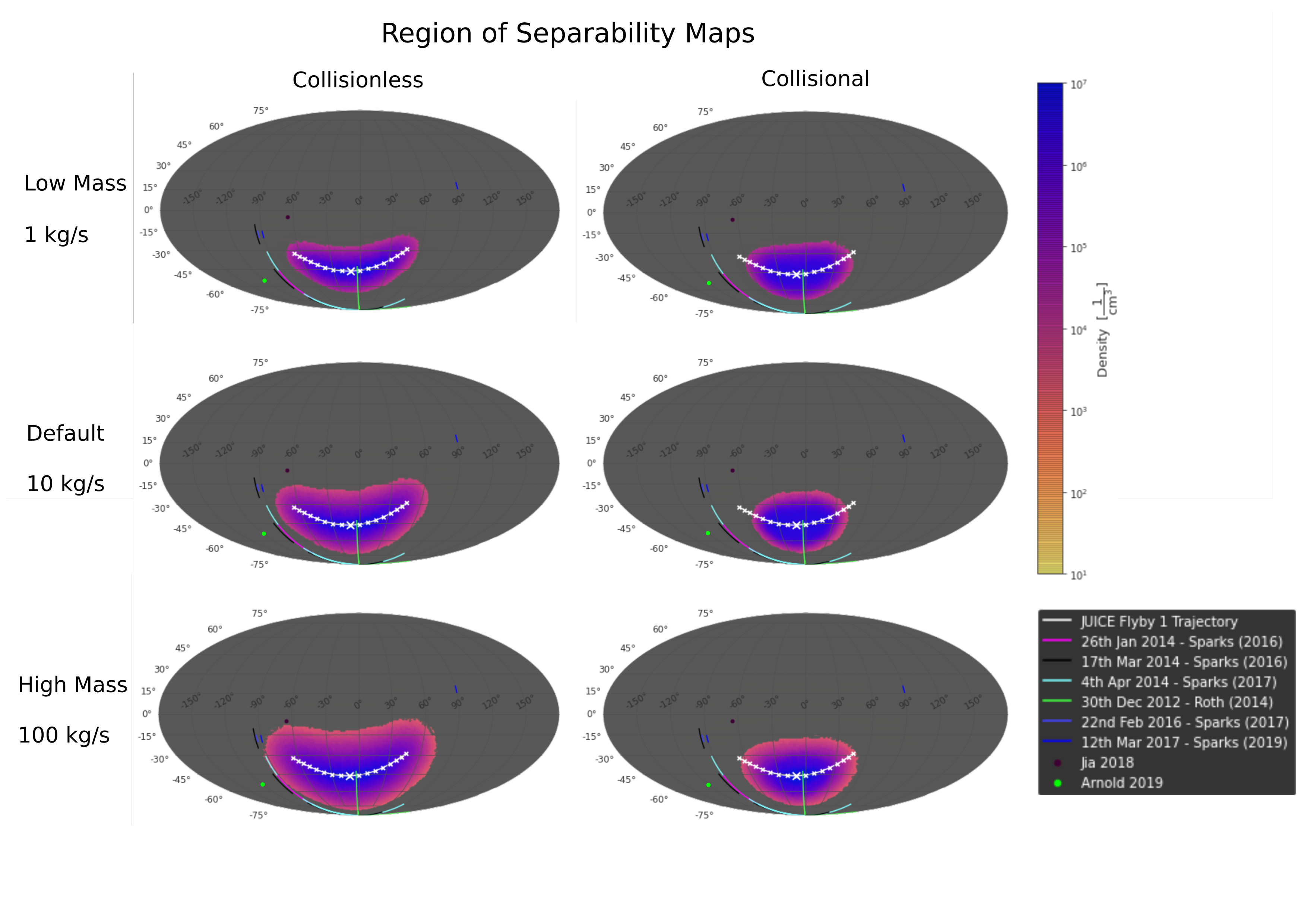}
    \caption{Region of separability maps for different plume models including the collisional and collisionless cases for low 1\unit{kg\,s^{-1}}, default 10\unit{kg\,s^{-1}}, and high 100\unit{kg\,s^{-1}} mass flux plumes. The region of separability is defined as the area over the surface of Europa where a plume located at that point would be separable from the atmosphere i.e. it exceeds the atmospheric threshold. This plot illustrates that the region of separability is reduced for a collisional plume model, for a low mass plume only by a small amount, but for a high mass plume the potential region of separability is reduced by a half.}
    \label{fig:appendix_ros}
\end{figure}

\end{document}